\documentclass[floats,prb,aps,showpacs]{revtex4}
\usepackage{graphicx}
\usepackage{epsfig}
\usepackage{bm}
\begin{document}

\title{Energy bands, conductance and thermoelectric power for ballistic
electrons in a nanowire with spin-orbit interaction}

\author{Godfrey Gumbs$^{1,4}$}
\email{ggumbs@hunter.cuny.edu}
\author{Antonios Balassis$^2$}
\email{balassis@fordham.edu}
\author{Danhong Huang$^3$}
\email{danhong.huang@kirtland.af.mil}
\address{$^1$Department of Physics and Astronomy, Hunter College of the
City University of New York, 695 Park Avenue, New York, NY 10065, USA}
\address{ $^2$Physics Department, Fordham University, 441 East Fordham Road,
       Bronx, NY 10458, USA}
\address{$^3$Air Force Research Laboratory, Space Vehicles
Directorate,\\
Kirtland Air Force Base, NM 87117, USA}
\address{$^4$Donostia International Physics Center (DIPC),
P. de Manuel Lardizabal, 4, 20018 San Sebasti\'an,
Basque Country, Spain }

\date{\today}

\begin{abstract}
We calculated the effects of spin-orbit interaction (SOI)  on the energy bands,
ballistic conductance ($G$) and the electron-diffusion thermoelectric power
($S_{\rm d}$)
of a nanowire by varying the temperature, electron density and width of the
wire.
The potential barriers at the edges of the wire are assumed to be very high. A
consequence of the boundary conditions used in this model is determined by the
energy band structure,
resulting in wider plateaus when the  electron density is increased
due to larger energy-level separation as the higher subbands are occupied by
electrons.
The nonlinear dependence of the transverse confinement on position with respect
to the well center
excludes the ``pole-like feature" in $G$ which is obtained when a harmonic
potential is
employed for confinement. At low temperature, $S_{\rm d}$ increases linearly
with $T$ but
deviates from the linear behavior for large values of $T$.
\end{abstract}

\pacs{ 73.23.Ad, 75.70.Tj,81.07.Gf}
\maketitle

\section{Introduction}
\label{sec1}

The electronic transport and photonic properties of a two-dimensional electron
gas
(2DEG) such as that found at a semiconductor heterojunction of GaAs/AlGaAs
have been the subject of interest and discussion for many years
now.\,\cite{davies}
Related physical properties of narrow quantum wires of 2DEG have also been the
subject
of experimental and theoretical investigations because of their potential
for device applications in the field of nanotechnology.\,\cite{pepper1,pepper2}
It is thus necessary to specify the model for the edge of a narrow quantum
wire.\,\cite{moroz,privman,gumbs1} Here, we analyze the role
played by the boundaries on the ballistic electron transport in a nanowire of
2DEG where
the Rashba spin-orbit interaction (SOI) is included. The role of SOI on
collective
properties of the 2DEG has been investigated.\,\cite{manvir,gumbs,go}
The quasi-one-dimensional channel may be formed by applying a negative bias to a
metal gate placed on the surface and depleting the 2DEG below it. Alternatively,
the quasi-one
dimensional channel may be made by etching all the way down to the active layer.In either way, it has been demonstrated that the width of the wire could
influence
the electron transport properties.\,\cite{pepper1,pepper2}
In the latter case, the edges are sharper and  may be appropriately modeled by
sharp and high boundary conditions. More recently, Brey and
Fertig\,\cite{fertig} have pointed
out that graphene nanoribbons have different collective plasmon dispersion
relations depending on whether they have armchair or zigzag edges.
\medskip

It is well established that the spin-orbit coupling is an essentially
relativistic effect:
an electron moving in an external electric field sees a magnetic field in its
rest frame.
In a semiconductor, the interaction causes an electron's spin to precess as it
moves through the
material, which is the basis of various proposed ``spintronic" devices. In
nano-structures,
quantum confinement can change the symmetry of the spin-orbit interaction. The
relativistic
motion of an electron is described by a Dirac equation. These effects combine to
form both
an electric dipole moment and the Thomas precession which is due to the
rotational
kinetic energy in the electric field.\,\cite{darwin,fisher} The two mechanisms
accidentally
have very close mathematical form and consequently combine in a very elegant
way.
The SOI Hamiltonian can be obtained from the Dirac equation by taking the
non-relativistic
limit up to terms quadratic in $v/c$. This limit can be achieved either by
expanding
the Dirac equation in powers of $v/c$ or by making use of the asymptotically
exact Foldy-Wouthousen transformation.\,\cite{foldy}
\medskip

\begin{figure}[p]
\begin{center}
\includegraphics[width=3.0in,height=3.0in]{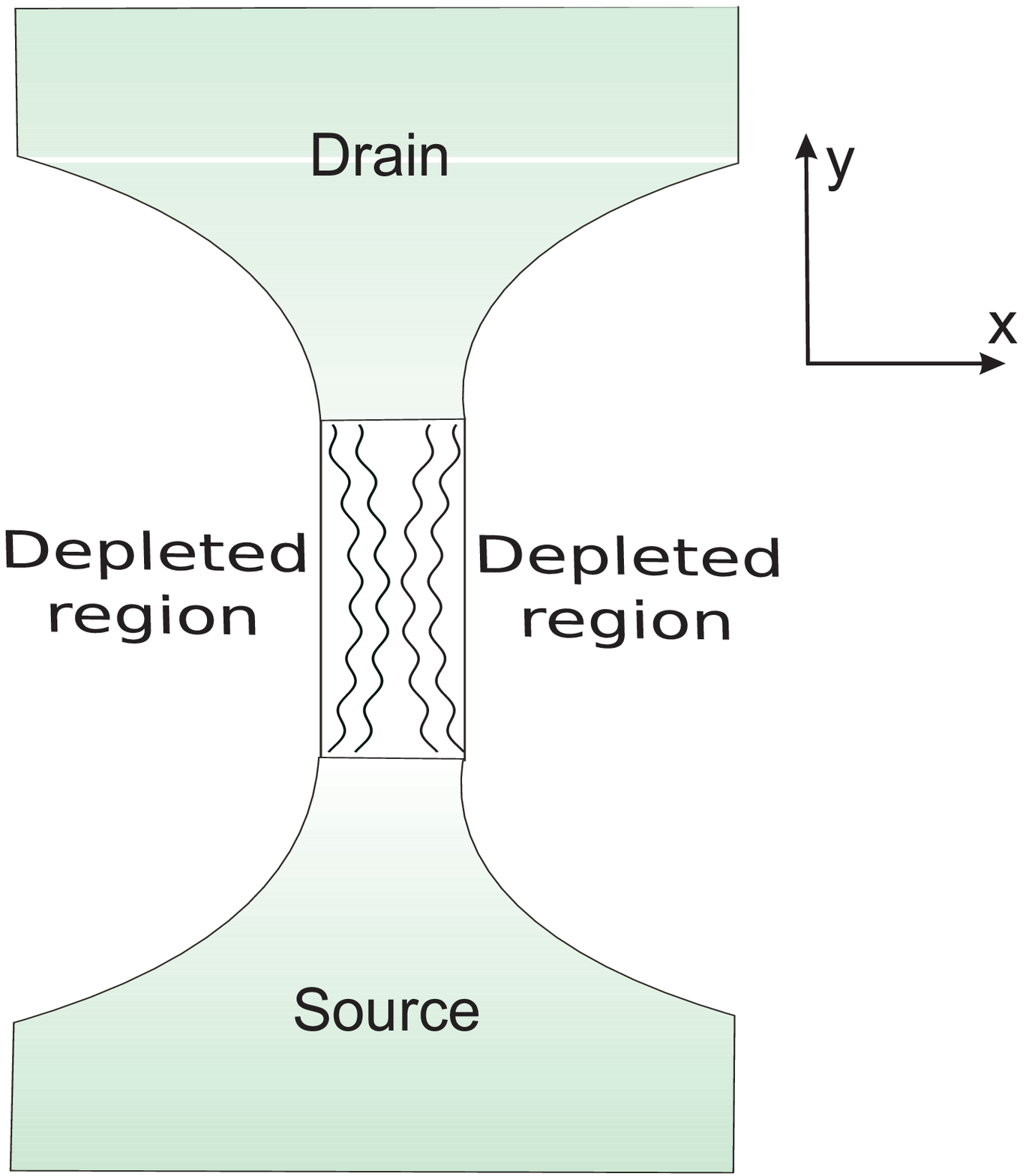}
\caption{(Color online) Schematic illustration of the nanowire of 2DEG between a
source and drain.}
\label{FIG:1}
\end{center}
\end{figure}

We include the effects due to edges through sharp and high potentials
at the boundaries. As a result,
we are not able to solve the Rashba SOI model Hamiltonian to obtain analytic
solutions
for the eigenenergies and eigenfunctions.  The reason for this is
due to the fact that  the solution manifestly contains quantum interference
effects from multiple scattering  off the edges. We solved the eigenvalue
problem numerically, obtaining the energies as a function of the wave vector
$k_y$ parallel to the edge of the nanowire shown schematically in
Fig.\,\ref{FIG:1}.  This model for the edges is different from that employed in
previous works.\,\cite{moroz,privman,gumbs1}
We calculate the ballistic conductance and the electron-diffusion thermoelectric
power
for this quasi-one-dimensional structure
by assuming that the length of the channel between source and drain is less
than the electron mean free path. In addition, we assume that the width
of the channel is of the order of  the de Broglie wavelength.\,\cite{Beenakker}
\medskip

The outline of the rest of this paper is as follows. In Sec.\,\ref{sec2},
we present our model for the wide quantum wire with spin-orbit coupling and
specified boundary conditions to simulate the effects arising from the edges of
the nanowire. We also present numerical results for the energy bands in
order to study the combined effect of the boundaries and the SOI.
Section\ \ref{sec3} is devoted to a brief description of the way in which
our calculations are done for the ballistic conductance and electron-diffusion
thermoelectric power when the energy bands
are symmetric with respect to the wave vector $k_y$ parallel to the edges
of the nanowire.\,\cite{LYO:2004} Numerical results and discussion of the
conductance
and the electron-diffusion thermoelectric power as functions of electron densityand temperature, for various wire widths and Rashba parameters, are given in
Sec.\,\ref{sec4}. A summary of our
results is presented in Sec.\,\ref{sec5}.

\section{Model for Energy Band Structure}
\label{sec2}

It is now well established that, spin-orbit coupling is an essentially
relativistic effect. The relativistic motion of an electron is described
by the Dirac equation that contains both effects (electric dipole and Thomas
precession) in the spin-orbit interaction and does so in a very elegant way
(see, e.g., textbooks\,\cite{num12,num13}). The SOI Hamiltonian can
be obtained from the Dirac equation by taking the non-relativistic limit of
the Dirac equation up to terms quadratic in $v/c$ inclusive. This limit can
be attained in two different ways: by direct expansion of the Dirac equation
in powers of $v/c$ and by the asymptotically exact Foldy-Wouthuysen
transformation\,\cite{foldy}. The Hamiltonian for an electron in the quadratic
[$O(v^2/c^2)$]
approximation is the sum

\begin{equation}
\tilde{{\cal H}}= \hat{\cal H}_{\rm SO}+\Delta \hat{H}\ ,
\label{approx9}
\end{equation}
where $\Delta \hat{H}$ is the free-particle  Hamiltonian  and

\begin{equation}
\hat{\cal H}_{\rm SO}=\frac{\hbar}{4m^2c^2}\left[\overrightarrow{\nabla}V({\bf
r})\times
\hat{{\bf p} } \right]\cdot \tensor{{\bf \sigma}}
\label{SOI-1}
\end{equation}
describes the SOI within the material and includes both  contributions to the
spin-orbit coupling from the electric dipole and the Thomas precession
(caused by the electric field) mechanisms. This result is {\em general\/}
since  it was derived from the Dirac equation,  an exact
relativistic equation for the electron, and includes all possible relativistic
effects, whatever might be  their kinetic source. When an electron gas at a
heterojunction is confined to the $xy$-plane so that the electrostatic potentialis spatially-uniform along the heterostructure interface and varies only along
the
$z$ axis, the Hamiltonian in Eq.\,(\ref{SOI-1}) contains just the contribution
arising from its confinement along the $z$ direction. For a
quasi-one-dimensional
structure, a second term must now be added to account for the extra local
confinement produced by the electric field within the $xy$-plane.

For quantum wires, the width of the potential well is comparable with the
spatial spread of
the electron wave functions in the $z$ direction. Therefore, in order to
determine
an effective electric field acting on electrons in the potential well, one
should
calculate an average of the electric field $E(z)$ over the range of the $z$
variable where the wave function is essentially finite. Consequently, one can
model the averaged electric field by a potential profile. In principle, all
potential
profiles can be classified in two ways. In the first case, the average of $E(z)$
is negligible although $E(z)$ itself may not be zero or even small. This applies
for
symmetric potentials, such as the square and parabolic quantum wells. However,
for asymmetric quantum wells, the average electric field is non-zero
in the direction perpendicular to the plane of the 2DEG and is called the
interface or quantum well electric field.  For experimentally achievable
semiconductor heterostructures, this field can be as high as $10^7$\,V/cm.
Therefore, from Eq.\,(\ref{SOI-1}), there should be an additional (compared
with the infinite 3D crystal) mechanism of spin-orbit coupling associated
with this field and is usually referred to as the Rashba SOI for quantum
wells.\,\cite{Rashba-paper} When we take into account that the quantum well
electric field is perpendicular to the heterojunction interface, the spin-orbit
Hamiltonian has a contribution which can be written for the Rashba coupling as

\begin{equation}
\hat{\cal H}_{\rm SO}^{(\alpha)}=\frac{\alpha_{\rm R}}{\hbar}\,
\left(\tensor{{\bf \sigma}} \times \hat{{\bf p}}\right)_z
\label{SOI-2}
\end{equation}
within the zero $z$-component (stationary situation, no electron transfer across
the interface).
The constant $\alpha_{\rm R}$ in Eq.\,(\ref{SOI-2}), which will be simply
denoted as $\alpha$ thereafter in this paper, includes universal constants from
Eq.\,(\ref{SOI-1}) and it is proportional to the the interface electric field.
The value of $\alpha$ determines the contribution of the Rashba spin-orbit
coupling to the total electron Hamiltonian. This constant may have values
running from
$(1 - 10)$\,meV$\cdot$\,nm.

\begin{figure}[p]
\begin{center}
\includegraphics[width=3.0in,height=2.5in]{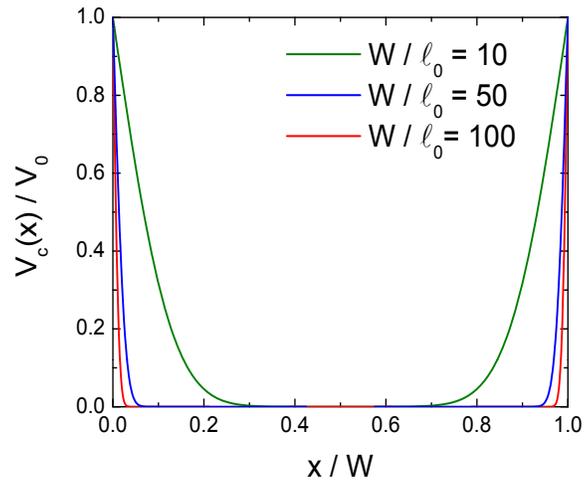}
\caption{(Color online) Plots of $V_{\rm c}(x)/V_0$, defined in
Eq.\,(\ref{error-function}), as a function of $x/{\cal W}$
for ${\cal W}/\ell_0=10$ (green curve), ${\cal W}/\ell_0=50$ (blue curve), and
${\cal W}/\ell_0=100$ (red curve).}
\label{f0}
\end{center}
\end{figure}

Within the single-band effective mass approximation\,\cite{num19,num20}, the
total Hamiltonian of a quasi-one-dimensional electron system (Q1DES) can be
written as

\begin{equation}
\hat{{\cal H}}=\frac{\hat{{\bf p}}^2}{2m^\ast}+V_{\rm c}({\bf r})+\hat{\cal
H}_{\rm SO}
\label{SOI-3}
\end{equation}
where the electron effective mass $m^\ast$
incorporates both the crystal lattice and interaction effects.
The form of the Hamiltonian derived from the relativistic $4\times 4$ Dirac
equation is similar to that which follows from the $8\times 8$ ${\bf k}\cdot
{\bf p}$
Hamiltonian\,\cite{referee}. Moroz
and Barnes\,\cite{moroz} chose the lateral confining potential $V_{\rm c}({\bf
r})$
as a parabola which would be appropriate for very narrow wires since
the electrons would be concentrated at the bottom of the potential.
Such narrow Q1DES are difficult to achieve experimentally. We are not aware
of any experimental evidence or measurement of the features arising
from the spin-orbit coupling resulting from the parabolic confining potential
employed by Moroz and Barnes \cite{moroz}. So, in this paper,
we explore the effects of lateral confinement in which the electrons are
essentially
free over a wide range except close to the edges where the potential
rises sharply to confine
them.  The in-plane electric field ${\bf E}_{\rm c}({\bf r})$ associated with
$V_{\rm c}({\bf r})$ is given by ${\bf E}_{\rm c(}{\bf r})=
-\overrightarrow{\nabla}V_{\rm c}({\bf r})$.
We assume that the SOI Hamiltonian in Eq.\,(\ref{SOI-3}) is formed by two
contributions:
$\hat{\cal H}_{\rm SO}=\hat{\cal H}_{\rm SO}^{(\alpha)}+\hat{\cal H}_{\rm
SO}^{(\beta)}$.
The first one, $\hat{\cal H}_{\rm SO}^{(\alpha)}$,
[in Eq.\,(\ref{SOI-2})] arises from the asymmetry of the quantum well, i.e.,
from
the Rashba mechanism\,\cite{Rashba-paper} for the spin-orbit coupling.
For convenience, in what follows we will refer to the Rashba mechanism of the
spin-orbit coupling as $\alpha$-coupling. If the lateral confinement is
sufficiently strong, for narrow and deep potentials or sharp and high potentialsat the edges, then the electric field associated with it may not be negligible
compared with the interface-induced (Rashba) field.  We use

\begin{equation}
V_{\rm c}(x)=V_0\left\{\mbox{erfc}\left(\frac{x}{\ell_0\sqrt{2}}  \right)
+ \mbox{erfc}\left(\frac{{\cal W}-x}{\ell_0\sqrt{2}}  \right)     \right\}
\label{error-function}
\end{equation}
for a conducting channel of width ${\cal W}$ with well depth $V_0$.  Here,
${\rm erfc}(x)$ is the complimentary error function. Plots of $V_{\rm c}(x)/V_0$as a function of $x/{\cal W}$ are shown in
Fig.\,\ref{f0} for three values of ${\cal W}/\ell_0$.
For this potential, the Hamiltonian (\ref{SOI-1})  gives a term

\begin{equation}
\hat{\cal H}_{\rm SO}^{(\beta)}=-i\beta\sigma_{\rm z}\left(\frac{{\cal
W}}{\ell_0}\right)\,\left\{\exp\left[-\frac{(x-{\cal W})^2}{2\ell_0^2}\right]-
\exp\left[-\frac{x^2}{2\ell_0^2}\right]\right\}\,\frac{\partial}{\partial y}
\equiv i\beta {\cal F}(x)\,\sigma_{\rm z}\,\frac{\partial}{\partial y}\ ,
\label{beta}
\end{equation}
where each  Gaussian has width $\ell_0$ at the edges $x=0$
and $x={\cal W}$. In Eq.\,(\ref{beta}),
${\cal F}(x)$ is related to the electric field due to
lateral confinement in the $x$ direction. Since $\ell_0\ll{\cal W}$
characterizes the
steepness of the  potentials at the  two edges, we are at liberty to use
a range of values of the ratio of these two lengths, keeping in mind that the
in-plane confinement must be appreciable if the $\beta$-term is to play a role.
Therefore, in most of our calculations, we use only one small value of
$\ell_0/{\cal W}$ to illustrate the effects arising from our model on the
conductance and thermoelectric power. We introduced
the parameter $\beta_0=\hbar^2V_0/(4\sqrt{2\pi}\,m^{\ast\ 2}c^2{\cal W})$,
which is expressed in terms of fundamental constants as well as $V_0$ and ${\cal
W}$.
The $\beta_0$ is another Rashba parameter due to the electric confinement along
the
$x$ direction, and it is simply denoted as $\beta$ thereafter in this paper.
Comparison of typical electric fields originating from the quantum
well and lateral confining potentials allows one to conclude that a reasonable
estimate\,\cite{moroz} for $\beta$ should be roughly 10\% of $\alpha$. The
$\beta$-SOI
term in Eq.\,(\ref{beta}) is asymmetric about the mid-plane $x={\cal W}/2$ and
varies quadratically with the displacement from either edge. In this
quasi-square well
potential, the electron wave functions slightly penetrate into the barrier
regions. However, we
only need energy levels for the calculations of ballistic transport electrons,
not the wave functions, if we assume electronic system is a spatially-uniform
quasi-one-dimensional one.
\medskip

The eigenfunctions for the nanowire have the form

\begin{equation}
\varphi({\bf r})= \frac{e^{ik_yy}}{\sqrt{L_y}}\,\left[\matrix{\psi_A(x)\cr
\psi_B(x) \cr}\right]\ .
\label{state}
\end{equation}
Since the nanowire is translationally invariant in the $y$-direction with
$k_y=(2\pi/L_y)\,n$,
where $L_y$ is a normalization length and $n=0,\,\pm1,\,\pm2,\,\cdots$, we must
solve for $\psi_A(x)$ and $\psi_B(x)$ in Eq.\,(\ref{state}) numerically
due to the presence of edges at $x=0$ and $x={\cal W}$. Substituting the
wave function in Eq.\,(\ref{state}) into
the Schr\"odinger equation, i.e. $\hat{{\cal H}}\varphi({\bf
r})=\varepsilon\varphi({\bf r})$ with
$\varepsilon$ being the eigenenergy, we obtain the two coupled equations

\begin{eqnarray}
& & -\frac{\hbar^2}{2m^\ast}\left(\frac{d^2}{dx^2}-k_y^2\right)\,\psi_A(x)
+\alpha\left(\frac{d}{dx}+k_y \right)\,\psi_B(x)-\beta k_y{\cal F}(x)\,\psi_A(x)=\varepsilon\,\psi_A(x)\ ,
\nonumber\\
& & -\frac{\hbar^2}{2m^\ast}\left(\frac{d^2}{dx^2}-k_y^2\right)\,\psi_B(x)
-\alpha\left(\frac{d}{dx}-k_y \right)\,\psi_A(x)+\beta k_y{\cal F}(x)\,\psi_B(x)=\varepsilon\,\psi_B(x)\ .
\label{coupled:eqs}
\end{eqnarray}
In the absence of any edges, we may simply set ${\cal F}(x)=0$ for a quantum
well, and get $\psi_A(x)={\cal A}\,e^{ik_xx}$
and $\psi_B(x)={\cal B}\,e^{ik_xx}$, where ${\cal A}$ and ${\cal B}$ are
independent of $x$, and $k_x$ is the electron wave number along the $x$
direction, which then yields a pair of simultaneous algebraic equations
for states $A$ and $B$. But, in the case when there exist edges, we have a pair
of
coupled differential equations to solve for $\psi_A$ and $\psi_B$ which may be
analyzed
when only $\beta$ is not zero and then when both Rashba parameters are non-zero.\medskip

Two parameters of interest are

\begin{equation}
\ell_\alpha=\hbar^2/2m^{\ast}\alpha\ , \ \ \ \
\ell_\beta=\hbar^2/2m^{\ast}\beta\ ,
\end{equation}
with three ratios

\begin{equation}
\tau_\alpha={\cal W}/\ell_\alpha\ , \ \ \ \ \tau_\beta={\cal W}/\ell_\beta\ ,
\ \ \ \ \tau_0={\cal W}/\ell_0\ .
\end{equation}
In our numerical calculations below, we will use three ratios to determine how
sharp the nanowire potential is and how strong the
Rashba parameters are.

\subsection{Energy Bands for $\alpha=0$}

\begin{figure}[hp]
\begin{center}
\includegraphics[width=3.5in,height=3.0in]{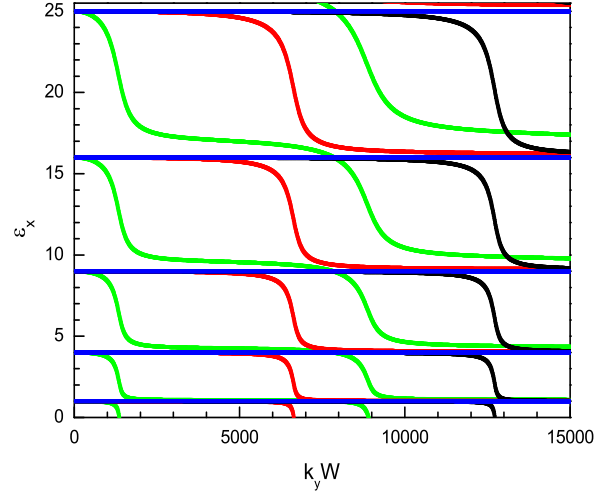}
\caption{(Color online) The transverse energy levels $\varepsilon_x$ scaled
with respect to $E_0$, the ground state energy
in the absence of any SOI, as a function of
$k_y{\cal W}$ for $\tau_{\beta}=0.1$ and $\tau_{\alpha}=0$.}
\label{FIG:2}
\end{center}
\end{figure}

When we set $\alpha=0$ in Eq.\,(\ref{coupled:eqs}), $\psi_A(x)$ and $\psi_B(x)$
are equal to each other ($k_y\to -k_y$) and are solutions of a Schr\"odinger
equation with a potential term present. When we solved the wave equations, we
imposed
the condition that the wave functions must vanish when either $x\ll 0$ or
$x\gg{\cal W}$ holds.
However, our calculations showed that the wave functions are negligible
on the two edges of the nanowire when the confining potential is deep and sharp.
The effect of the potential depends on $k_y$,
leading to a dependence of the transverse energy
$\varepsilon_x=\varepsilon-\hbar^2k_y^2/2m^\ast$ on the longitudinal wave number
$k_y$.
If we set  $\alpha$  equal to zero in Eq.\,(\ref{coupled:eqs}),
the solutions are approximately those for a quasi-square well when $\ell_0/{\cal
W}$ is chosen small
and the potential barriers are high (see Fig.\,\ref{f0}). In this case, the
transverse energy eigenvalues are approximately given by $E_n= n^2E_0$, where
$n=1,\,2,\,\cdots$ and $E_0\equiv\pi^2\hbar^2/(2m^{\ast}{\cal W}^2)$.
In Fig.\,\ref{FIG:2}, we present the energy bands for the calculated transverse
energy
$\varepsilon_x=\varepsilon-\hbar^2k_y^2/2m^\ast$ in units of $E_0$
when the electric field-induced Rashba SOI parameter $\alpha$ is set equal to
zero so
that only the effect from the $\beta$-term is included. The two equations
coincide
and, of course, there is no effect from the Rashba term on $\varepsilon_x$ when
$k_y=0$, and the eigenvalues are equal to those of the square well with high
barriers
corresponding to $\ell_0/{\cal W}\ll 1$. However, as $k_y{\cal W}$ is
increased, $\varepsilon_x$ decreases linearly as a function of $k_y$
before its first drop. On the other hand, the dependence of the transverse
energy has been shown to be a strong non-linear function of $k_y$ at its drop.
As a matter of fact, the levels anti-cross at a value
of $k_y{\cal W}$ which is determined by the chosen value of $\tau_\beta$
and $\ell_0/{\cal W}$.

The displacing effect of the $\beta$-coupling on the eigenstates is reminiscent
of the
role played by magnetic field on the eigenenergies of a quasi-one-dimensional
electron gas with harmonic confinement. However, the anti-crossing seems to be a
unique
property of the square barrier model with high potentials at the edges
since it was not reported by Moroz and Barnes\,\cite{moroz} for a parabolic
confinement.
Figure \ \ref{FIG:2} shows that when $\ell_0/{\cal W}\ll 1$, it does not matter
what value is chosen because the energy
level dependence on $k_y$ remains the same. When $k_y{\cal W}\gg 1$, the
influence on the $k_y$ dispersion can be seen. However, this
part of the energy spectrum makes a negligible contribution to the transport
and thermoelectric power.

\subsection{Energy Bands for $\alpha\neq 0$}

\begin{figure}[]
\begin{center}
\includegraphics[width=3.0in,height=2.5in]{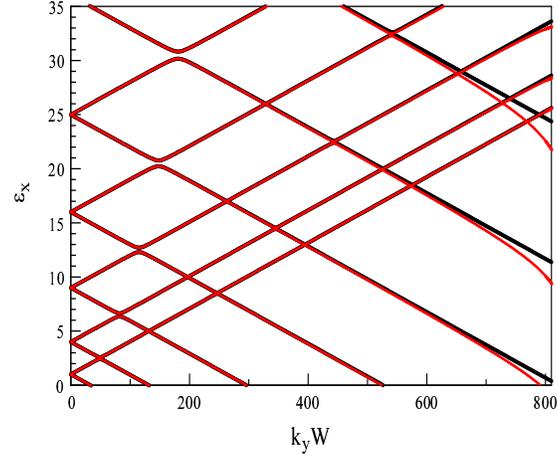}
\caption{(Color online) The transverse energy levels $\varepsilon_x$ scaled
with respect to $E_0$ as a function of
$k_y{\cal W}$ with $\tau_{\alpha}=0.3$ and $\tau_0=10^3$ for
$\tau_{\beta}=0$ (black curve) and $\tau_{\beta}=1.5$ (red curve).}
\label{FIG:3}
\end{center}
\end{figure}

In Fig.\,\ref{FIG:3}, we present the transverse energies $\varepsilon_x$ in
units of $E_0$ as functions of $k_y{\cal W}$ for a non-zero value of $\alpha$.
The plots compare the results for two chosen values of $\tau_{\beta}=0$ (black
curve) and $\tau_{\beta}=10$ (red curve). The energy bands for the infinite 2DEG
at $k_x=0$ due only to the $\alpha$-coupling (without transverse confinement)
consist of a pair of spin-split upward-curved parabolic-like
energy dispersions which are displaced in $k_y$-space and degenerate at $k_y=0$,
and become split as $k_y$ increases.
In the presence of edges for a nanowire, there is a discrete set of eigenstates
for a chosen value of $k_y$.
These energy subbands then anti-cross. This anti-crossing effect increases as
the
value of $\beta$ is increased, as demonstrated by comparing the results of
Fig.\,\ref{FIG:3}. These results are qualitatively in agreement
with Fig.\,3 in Ref.\,[\onlinecite{moroz}] (see also Ref.\
[\onlinecite{gumbs1}])
The difference is that for reasonable
values of $\tau_{\beta}$ and $\tau_{\alpha}$, the scaled $k_y$ becomes larger in
comparison with
a parabolic confinement for one to see the anti-crossing behavior.
\medskip

Figures\ \ref{FIG:4} show plots of total energy levels $\varepsilon$ with
$\tau_0=10^3$ as functions of $k_y{\cal W}$ for $\tau_{\alpha}=5.0$,
$\tau_{\beta}=1.0$ (upper-left panel) and
$\tau_{\alpha}=10.0$, $\tau_{\beta}=1.0$ (upper-right panel), respectively. For
the sake of comparison, the plot for $\tau_{\alpha}=0$, $\tau_{\beta}=2.0$, and
$\tau_0=10^3$ (lower panel) is also presented in this figure.
From Figs.\,\ref{FIG:4}, we easily find that the energy dispersion is symmetric
with respect to $k_y=0$. Our results (two upper panels) further show that,
as expected from Eq.\,(\ref{coupled:eqs}), the $\alpha$-term has an effect when
$k_y\neq 0$
on the unperturbed energy eigenvalues in the absence of any Rashba SOI. As
$k_y$ is increased, the SOI lifts the degeneracy of the energy subbands, as
shown by a pair of solid and dashed curves degenerated at $k_y=0$ in the upper
two panels. We further observed
that when the $\alpha$-coupling is weak (upper-left panel), each branch of the
energy
curves, except for the first branch, has only {\em one\/} local minimum and this
occurs at $k_y=0$. As
$\alpha$ is increased (upper-right panel), two local minima develop
symmetrically on
either side of $k_y=0$ with a local maximum at $k_y=0$ for each branch.

\section{Model for Ballistic Charge Transport}
\label{sec3}

In this section, we briefly outline our method of calculation for the ballistic
conductance and electron-diffusion thermoelectric power. In our case, the energy
bands are symmetric with respect to the wave numbers $\pm k_y$. For the
symmetric bands, energy dispersion $\varepsilon_{j,k_y}$, the Fermi function
$f_0(\varepsilon_{j,k_y})$, and the group velocity $v_{j,k_y}$ satisfy the
relations: $\varepsilon_{j,k_y}=\varepsilon_{j,-k_y}$,
$f_0(\varepsilon_{j,k_y})=f_0(\varepsilon_{j,-k_y})$, and
$v_{j,k_y}=-v_{j,-k_y}$. Therefore, one can write the following
equation\,\cite{LYO:2004} in a
form which includes only positive values of the wave number for the ballistic
heat (${\cal Q}^{(1)}$) and charge (${\cal Q}^{(0)}$) currents, i.e.,

\begin{eqnarray}
{\cal Q}^{(\ell)}=\frac{eV_{\rm
b}(-e)^{1-\ell}}{\pi}\,\sum_j\,\left(\int_{\varepsilon_{j,k_0}}^{\varepsilon_{j,k_1}}+\int_{\varepsilon_{j,k_1}}^{\varepsilon_{j,k_2}}+\cdots+\int_{\varepsilon_{j,k_N}}^{\infty}\right)\,
{\rm
sgn}(v_{j,k_y})\,\left(\varepsilon_{j,k_y}-\mu\right)^{\ell}\,\left[\frac{\partial
f_0(\varepsilon_{j,k_y})}{\partial\varepsilon_{j,k_y}}\right]\,
d\varepsilon_{j,k_y}\ ,
\label{basic}
\end{eqnarray}
where $\ell=0,1$, $V_{\rm b}$ is the bias voltage between the source and drain
electrodes, ${\rm sgn}(x)$ is the sign function, $k_0=0$, and $\mu$ is the
chemical potential. In Eq.\,(\ref{basic}), the whole energy integration
performed over the range $0\leq k_y<\infty$ is divided into the sum of many
sub-integrations between two successive extremum points $\varepsilon_{j,k_n}$
for $0\leq n\leq N$, and $\varepsilon_{j,k_N}$ is the last minimum point. For
each sub-integration over $k_y$, $\varepsilon_{j,k_y}$ is a monotonic function.
In addition, each sub-integration in Eq.\,(\ref{basic}) can be calculated
analytically, leading to the following expression for electron-diffusion
thermoelectric power

\begin{equation}
S_{\rm d}=\frac{{\cal Q}^{(1)}}{T{\cal Q}^{(0)}}=-\frac{k_{\rm
B}}{eg}\,\sum_{j,n}\,C_{j,n}\,\left[\beta\left(\varepsilon_{j,k_n}-\mu\right)f_0(\varepsilon_{j,k_n})+\ln\left(e^{\beta(\mu-\varepsilon_{j,k_n})}+1\right)\right]\
,
\label{thermo}
\end{equation}
where $T$ is the temperature, $\beta=1/k_{\rm B}T$, and the dimensionless
conductance $g$ is given by

\begin{equation}
g=\sum_{j,n}\,C_{j,n}\,f_0(\varepsilon_{j,k_n})\ .
\label{fermi}
\end{equation}
Physically, the quantity $g$ defined in Eq.\,(\ref{fermi}) represents the number
of pairs of the Fermi points at $T=0$\,K. In Eqs.\,(\ref{thermo}) and
(\ref{fermi}), the summations over $n$ are for all the energy-extremum points on
each $j$th spin-split subband in the range $0\leq k_y<\infty$. The quantity
$\varepsilon_{j,k_n}$ is the energy at the extremum point $k_y=k_{j,n}$. For a
given $j$th spin-split subband, $C_{j,n}=1$ (or $C_{j,n}=-1$) for a local energy
minimum (maximum) point. The physical conductance $G$ is related to $g$ for
spin-split subbands through

\begin{equation}
G=\left(\frac{e^2}{h}\right)g\ .
\label{cond}
\end{equation}

\section{Numerical Results of Charge Transport}
\label{sec4}

\begin{figure}[p]
\includegraphics[width=3.0in,height=2.5in]{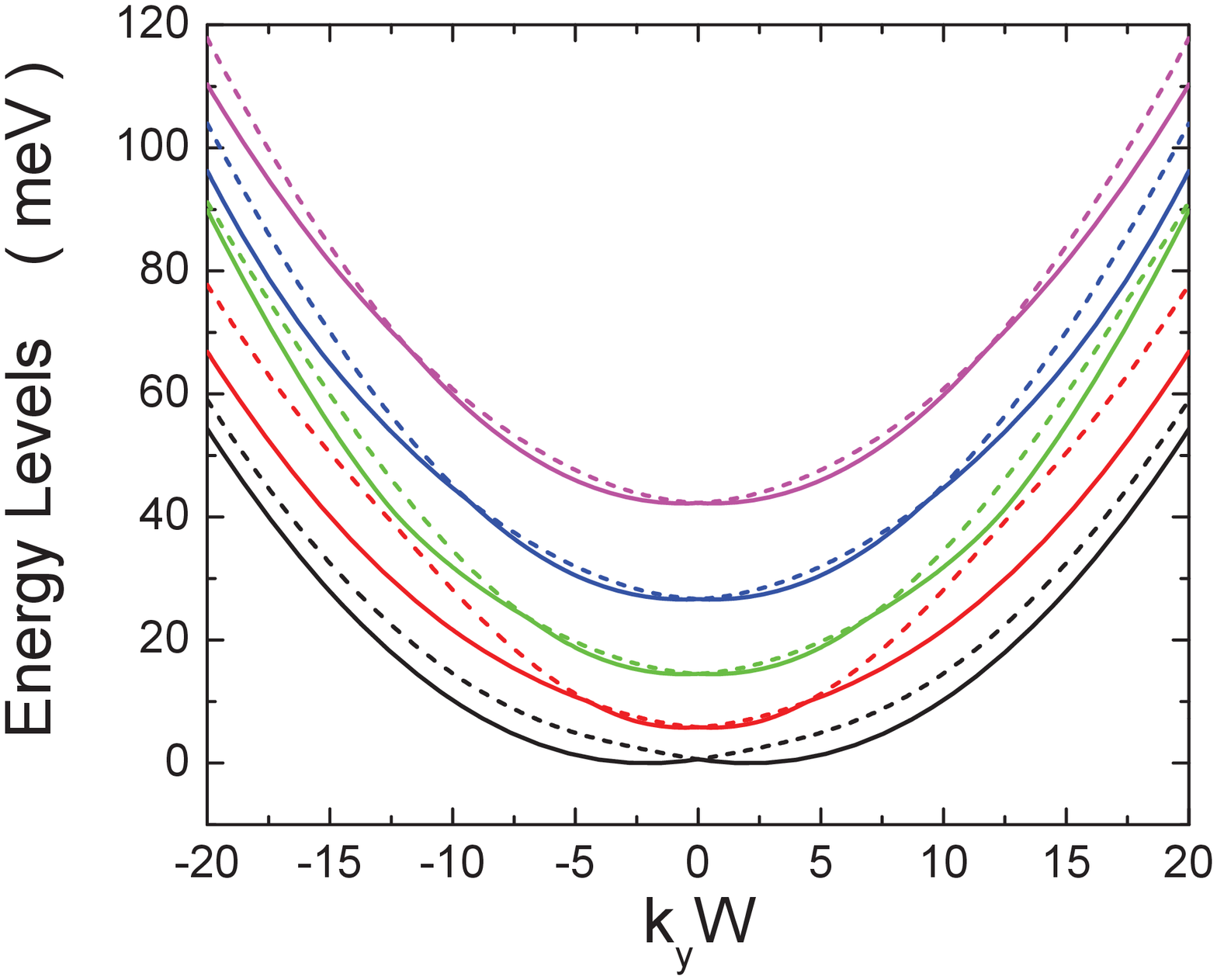}
\includegraphics[width=3.0in,height=2.5in]{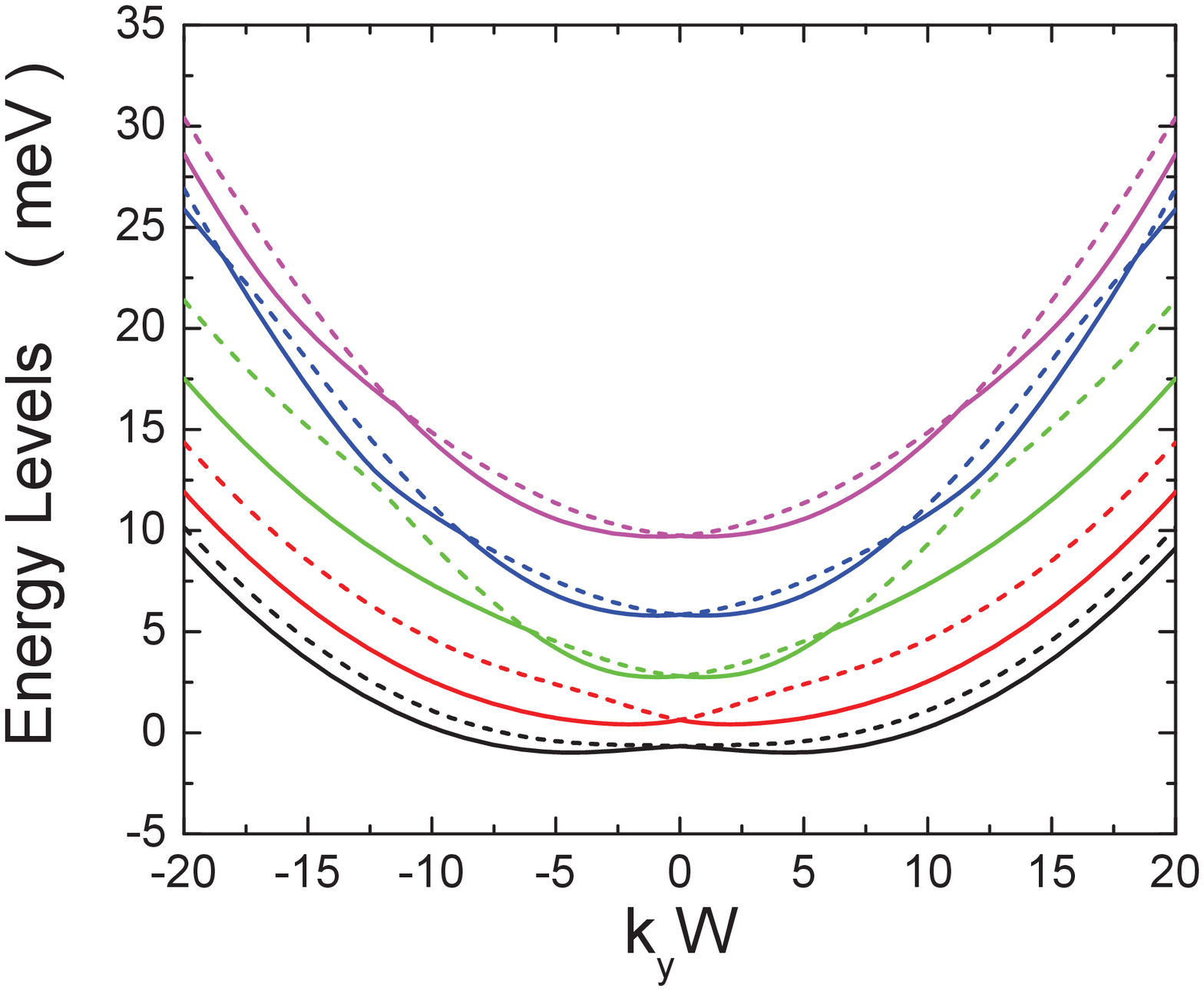}
\includegraphics[width=3.0in,height=2.5in]{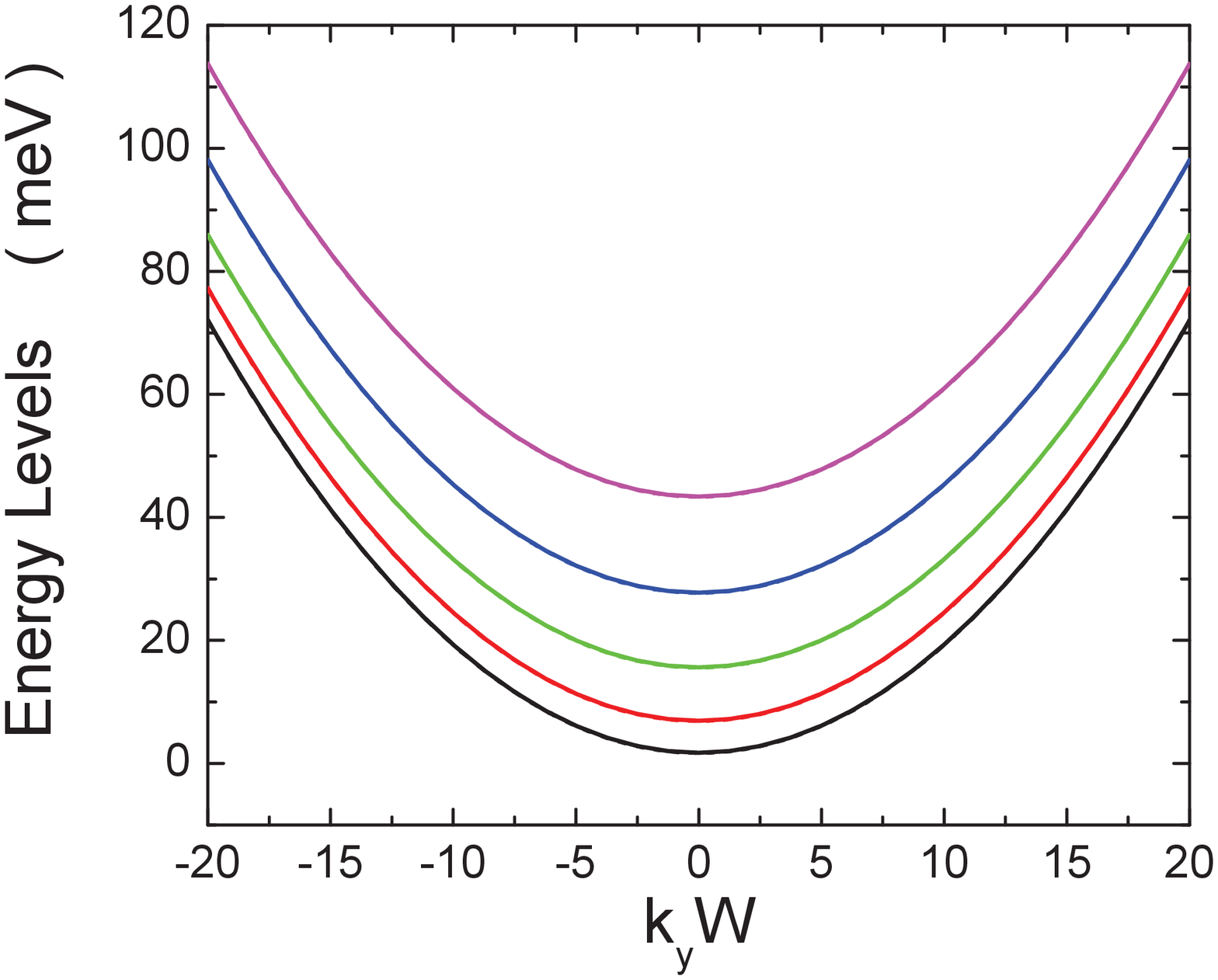}
\caption{(Color online) Plots of total energy levels $\varepsilon$
as functions of scaled wave number, $k_y{\cal W}$, for $\tau_{\alpha}=5.0$ and
$\tau_{\beta}=1.0$ (upper-left panel) and $\tau_{\alpha}=10.0$ and
$\tau_{\beta}=1.0$ (upper-right panel), respectively. Solid and dashed curves
represent a pair of spin-split subbands in a branch.
Here, the plot of spin-degenerated energy levels $\varepsilon$ for
$\tau_{\alpha}=0$, $\tau_{\beta}=2.0$ (lower panel) is also included for the
comparison. In our calculations, $\tau_0=10^3$ is chosen for these three
figures.}
\label{FIG:4}
\end{figure}

\begin{figure}[p]
\begin{center}
\includegraphics[width=4.0in,height=3.0in]{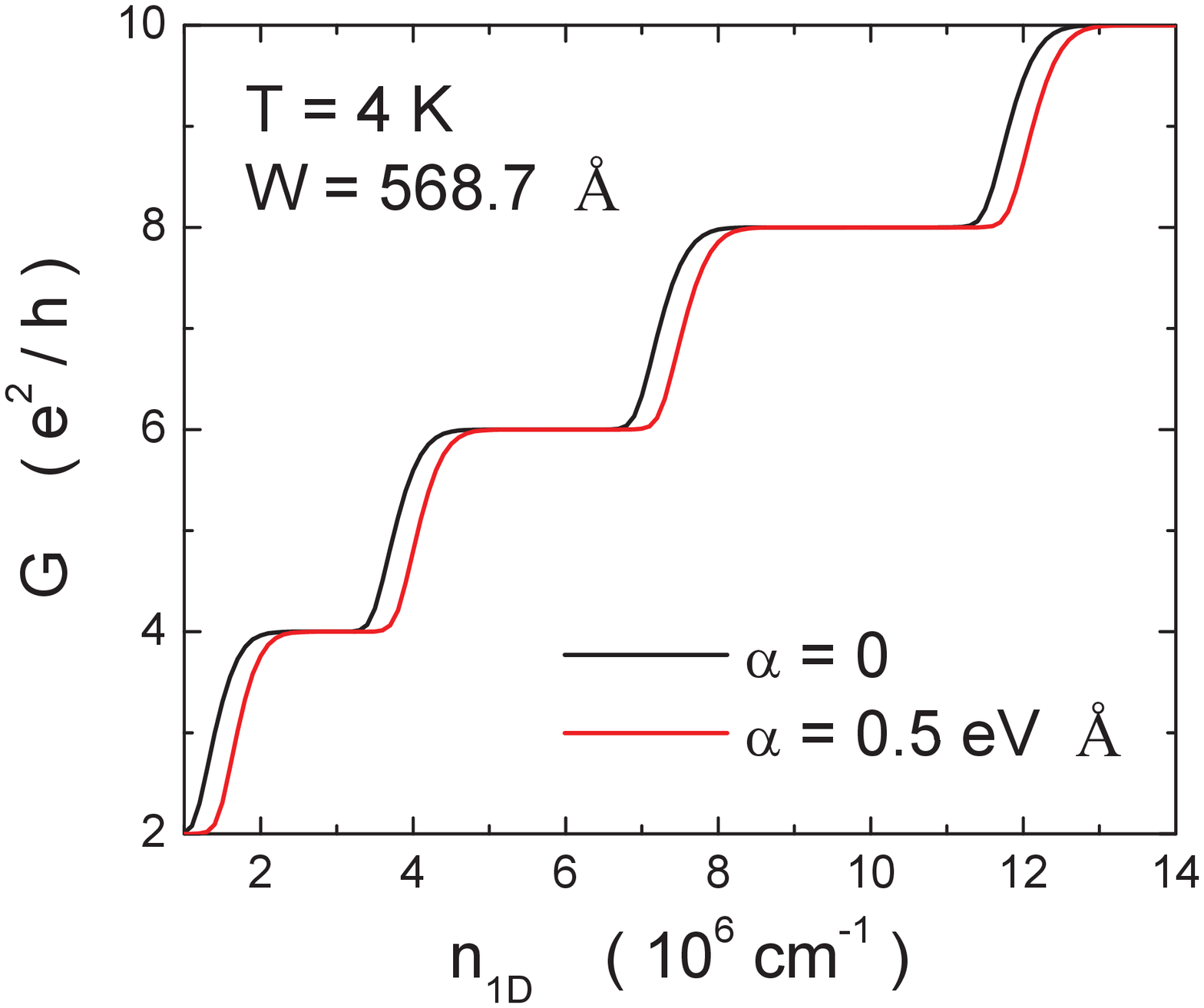}
\includegraphics[width=4.0in,height=3.0in]{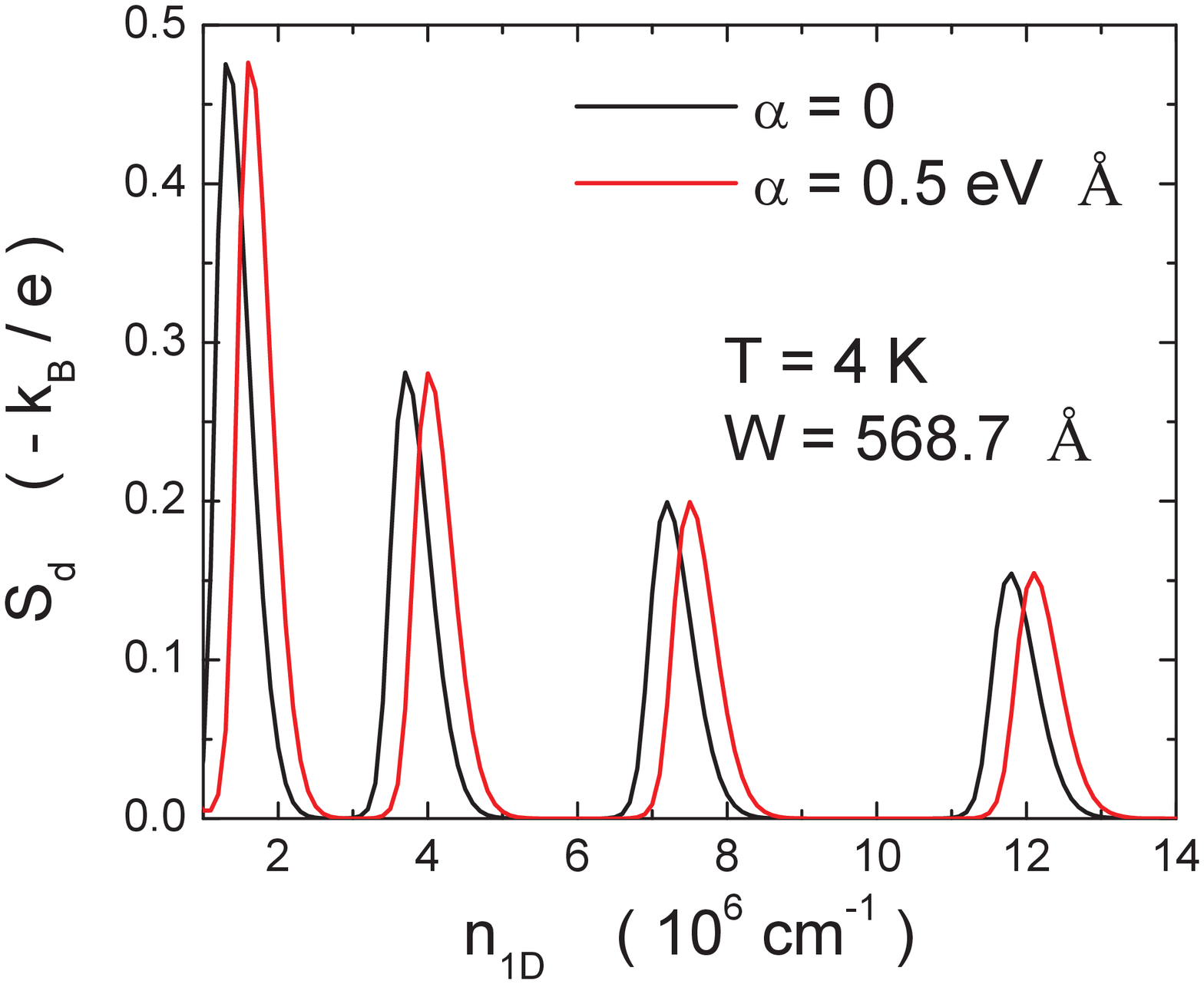}
\caption{(Color online) Comparisons of the conductance $G$ (upper panel), as
well as the electron-diffusion thermoelectric power $S_{\rm d}$ (lower panel),
as a function of the electron density $n_{\rm 1D}$ with a wire width ${\cal
W}=568.7$\,\AA\ and a temperature $T=4$\,K for $\alpha=0$ (black curves) and
$\alpha=0.5$\,eV$\cdot$\AA\ (red curves), respectively. Here, $\tau_{\beta}=1.0$
and $\tau_0=10^3$ are chosen for the calculations in these two figures.}
\label{FIG:5}
\end{center}
\end{figure}

In Figs.\,\ref{FIG:5}, we have displayed comparisons of modified electron
density ($n_{\rm 1D}$) dependence of the ballistic conductance ($G$) and the
electron-diffusion thermoelectric power ($S_{\rm d}$) by the $\alpha$-term in
the SOI when $T=4$\,K and ${\cal W}=568.7$\ \AA. From the upper panel of
Fig.\,\ref{FIG:5}, we find that, as $\alpha=0$ (black curve), a number of steps
in $G$ show up as a result of successive populations of more and more
spin-degenerate subbands, shown in the lower panel of Fig.\,\ref{FIG:4}. In
addition, the observed plateau becomes wider and wider as higher and higher
subbands are occupied by electrons due to increased energy-level separation,
resulting from the high potential barriers at the two edges. The
finite-temperature effect can easily be seen from the smoothed steps in this
figure. As $\alpha$ is increased to $0.5$\,eV$\cdot$\AA\ (red curve), the steps
are rightward shifted to higher electron densities due to an enhanced
density-of-states from the flattened subband dispersion curves by SOI, as seen
from the upper-left panel of Fig.\,\ref{FIG:4}. However, the step sharpness
remains constant in this case. Furthermore, there exists no ``pole-like
feature"\,\cite{moroz} in this figure, which can be traced back to the absence
of spike-like feature in the subband dispersion curves, leading to additional
local energy minimum/maximum points. The suppressed spike-like feature in the
subband dispersion curves can be explained by a nonlinear $x$ dependence near
the center ($x={\cal W}/2$) of a transverse symmetric potential well with a
large $\beta$ value in our model for wide quantum wires, instead of a linear $x$
dependence close to the center of the confining potential in the model proposed
by Moroz and Barnes\,\cite{moroz} for narrow quantum wires. We also see sharp
peaks in $S_{\rm d}$ from the lower panel of Fig.\,\ref{FIG:5} as $\alpha=0$
(black curve), corresponding to the steps in $G$, which again comes from
successive population of spin-degenerate subbands with increased $n_{\rm
1D}$.\,\cite{huang1} The center of a plateau in $G$ aligns with the minimum
of $S_{\rm d}$ between two peaks.\,\cite{huang1} The peaks (red curve) are
rightward shifted accordingly in electron density when a finite value of
$\alpha$ is assumed.
\medskip

\begin{figure}[p]
\begin{center}
\includegraphics[width=4.0in,height=3.0in]{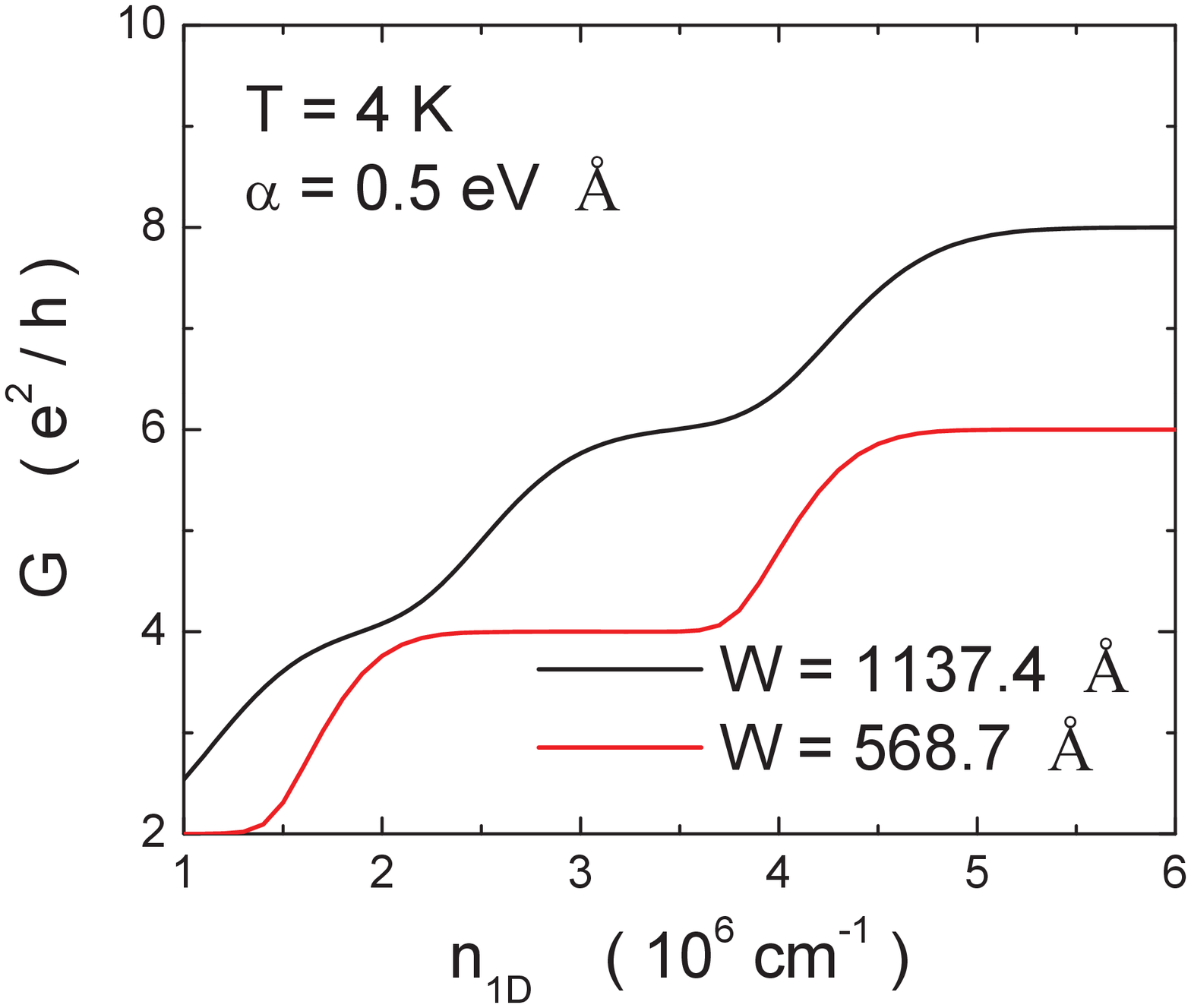}
\includegraphics[width=4.0in,height=3.0in]{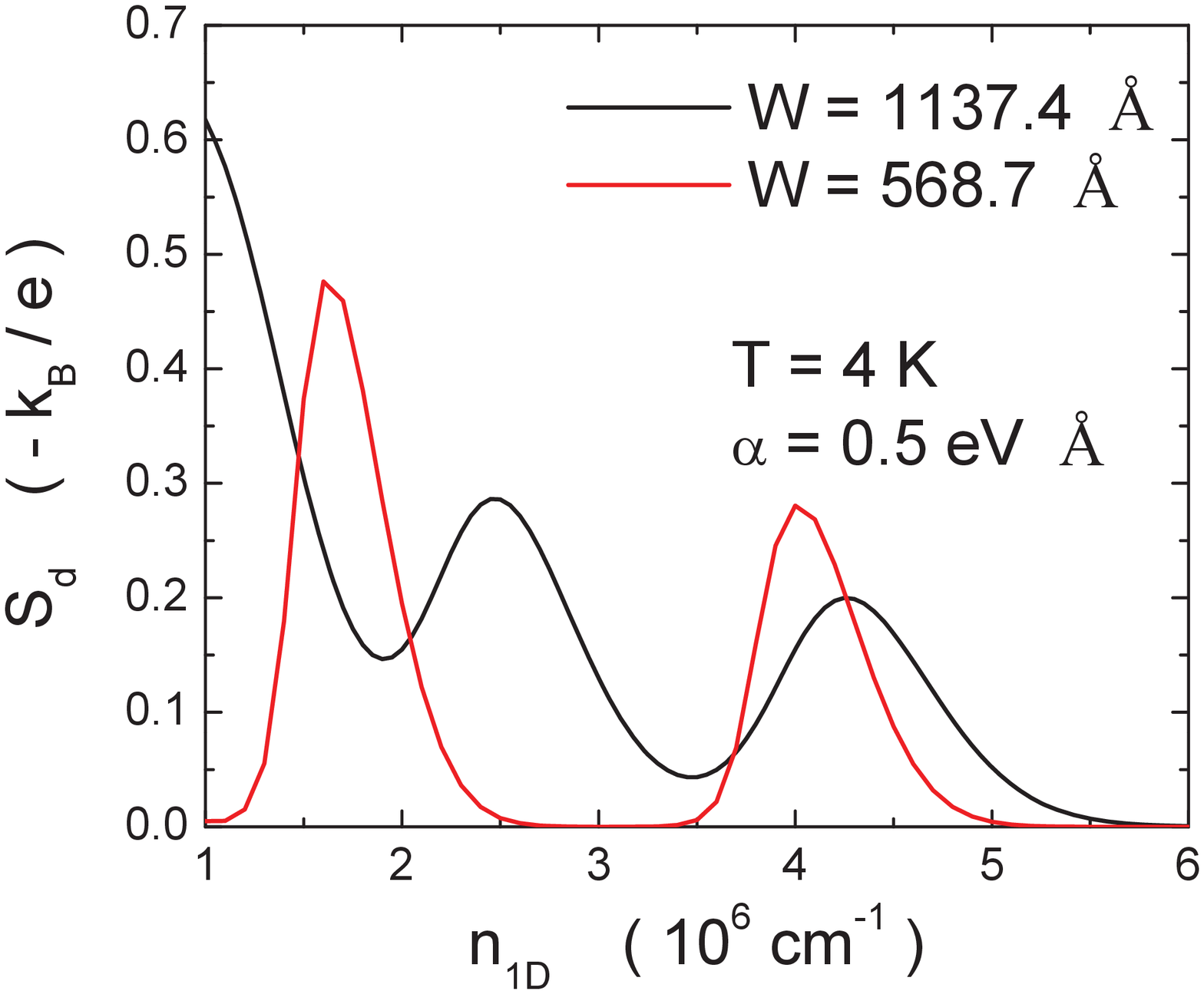}
\caption{(Color online) Comparisons of $G$ (upper panel) and $S_{\rm d}$ (lower
panel) as a function of $n_{\rm 1D}$ at $T=4$\,K and with
$\alpha=0.5$\,eV$\cdot$\AA\ for ${\cal W}=1137.4$\,\AA\ (black curves) and
${\cal W}=568.7$\ \AA\ (red curves), separately. Here, $\tau_{\beta}=1.0$ and
$\tau_0=10^3$ are chosen for the calculations in these two figures.}
\label{FIG:6}
\end{center}
\end{figure}

We have compared in Figs.\,\ref{FIG:6} the results of $G$ and $S_{\rm d}$ for
two values of wire width ${\cal W}$ at $T=4$\,K and $\alpha=0.5$\,eV$\cdot$\AA.
We find from the upper panel of Fig.\,\ref{FIG:6} that, as ${\cal W}$ decreases
from $1137.4$\ \AA\ (black curve) to $568.7$\ \AA\ (red curves), the steps in
$G$ are leftward shifted in electron density, and meanwhile, the steps become
sharpened. The step shifting is a result of the reduction of SOI effect due to a
smaller value for $\tau_{\alpha}$ (proportional to $\alpha{\cal W}$) with a
fixed value of $\alpha$, by comparing the upper-right panel with the upper-left
panel of Fig.\,\ref{FIG:4}. This leads to a leftward shift in steps for the same
reason given for the upper panel of Fig.\,\ref{FIG:5}. The step sharpening, on
the other hand, comes from the significantly increased subband separation
(proportional to $1/{\cal W}^2$), which effectively suppresses the
thermal-population effect on $G$ for smoothing out the conductance steps. The
shifting of steps in $G$ with ${\cal W}$ is also reflected in $S_{\rm d}$, as
shown in the lower panel of Fig.\,\ref{FIG:6}. The peaks of $S_{\rm d}$ get
sharpened due to the suppression of $S_{\rm d}$ in the density region
corresponding to the widened plateaus of $G$.
\medskip

\begin{figure}[p]
\includegraphics[width=7.0in,height=3.0in]{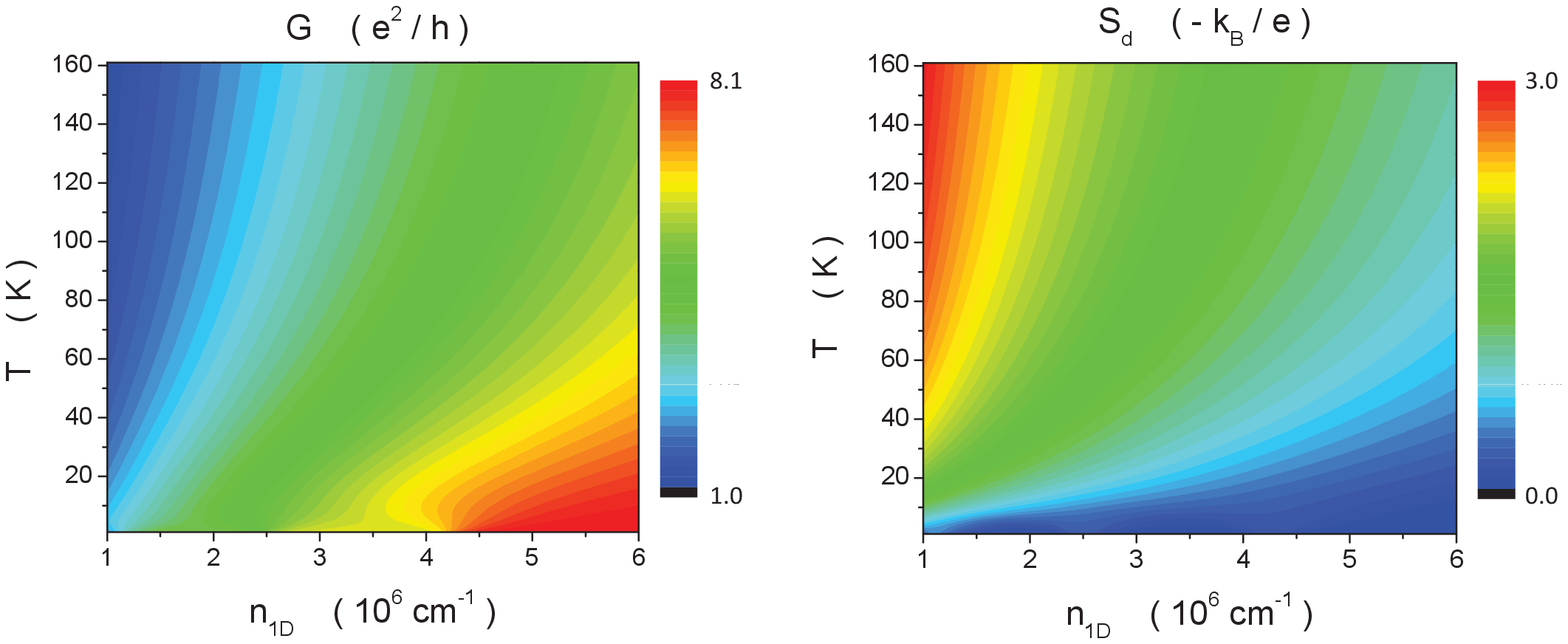}
\includegraphics[width=3.5in,height=2.5in]{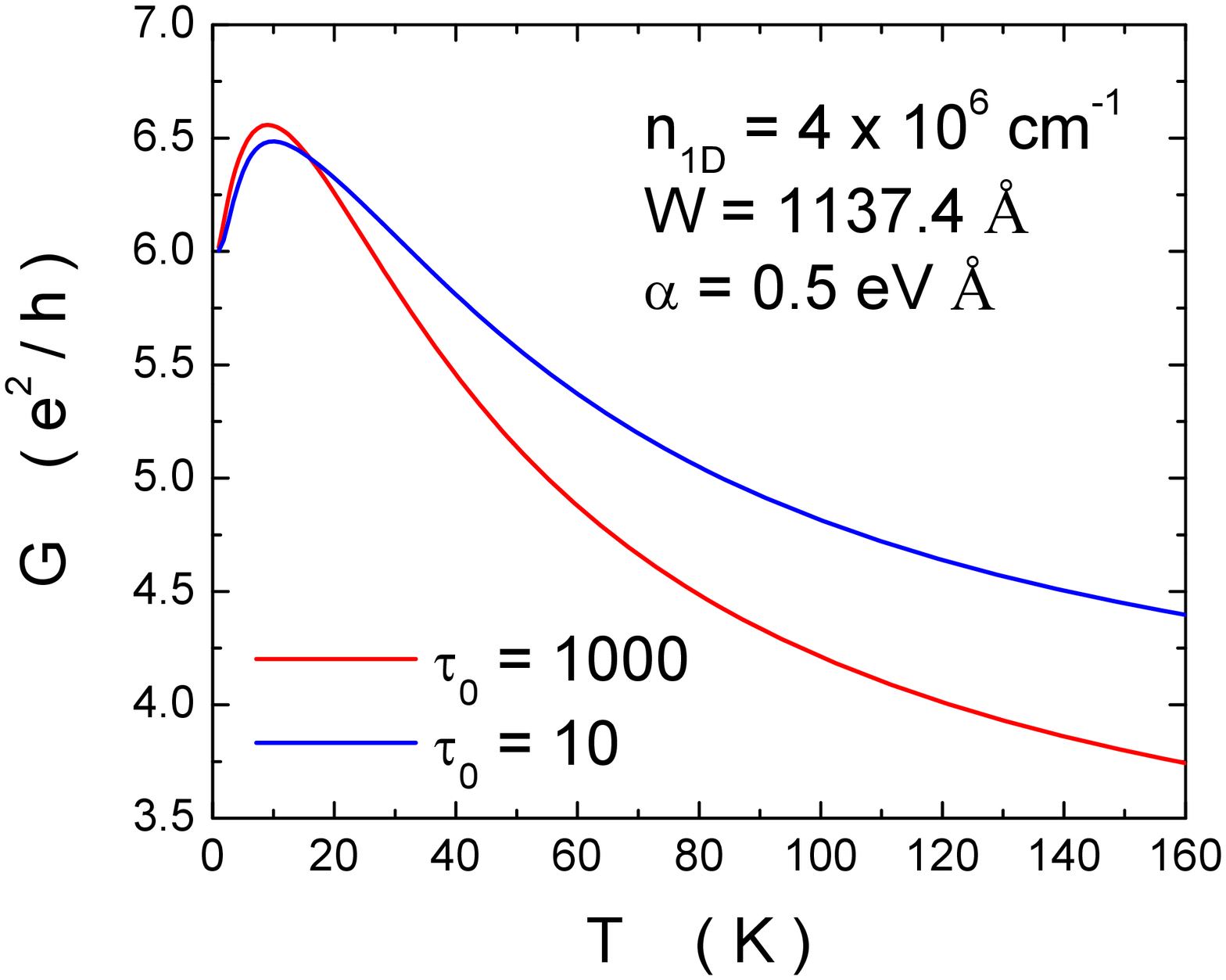}
\includegraphics[width=3.5in,height=2.5in]{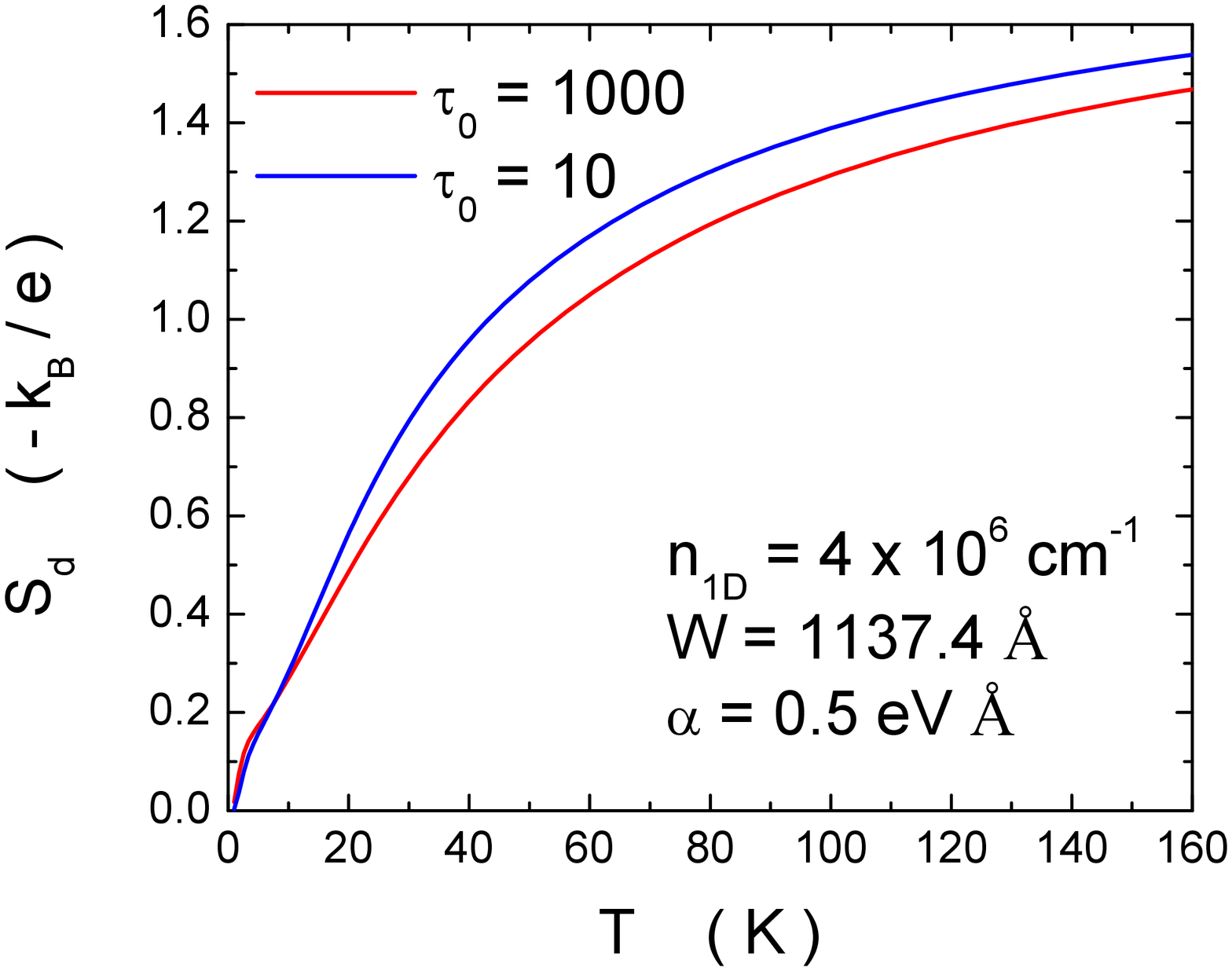}
\caption{(Color online) Contour plots of $G$ (upper left) and $S_{\rm d}$ (upper
right)
as functions of both $T$ and $n_{\rm 1D}$ with $\alpha=0.5$\,eV$\cdot$\AA\ and
${\cal W}=1137.4$\,\AA. Here, $\tau_0=10^3$ is chosen for the upper two Contour
plots. The plots of $G$ (lower left) and $S_{\rm d}$ (lower right) are also
shown in this figure as a function of $T$
for $n_{\rm 1D}=4\times 10^6$\,cm$^{-1}$ with $\tau_0=10^3$ (red curves) and
$\tau_0=10$ (blue curves).}
\label{FIG:7}
\end{figure}

In order to achieve an overview for the variations of $G$ and $S_{\rm d}$ with
both $T$ and $n_{\rm 1D}$, we present two contour plots of these quantities,
respectively, in Figs.\,\ref{FIG:7} with ${\cal W}=1137.4$\,\AA\ and
$\alpha=0.5$\,eV$\cdot$\AA. From the upper-left panel of Fig.\,\ref{FIG:7}, we
find that $G$ decreases with $T$, but increases with $n_{\rm 1D}$ in general.
The increase of $G$ with $n_{\rm 1D}$ is a direct result of opening more
conduction channels, i.e. more populated subbands, for ballistically-transported
electrons. The reduction of $G$ with $T$ is a consequence of the dramatic
decrease of the chemical potential with $T$ for a fixed $n_{\rm 1D}$ (not
shown), and then, the decrease of the Fermi function in Eq.\,(\ref{fermi}) for
$T>10$\,K. However, $G$ does increase with $T$ at $n_{\rm 1D}=4\times
10^6$\,cm$^{-1}$ within the range $T\leq 10$\,K, as shown in the lower-left
panel of Fig.\,\ref{FIG:7}, because of the anomalous increase of the chemical
potential with $T$ in this temperature range whenever the Fermi energy at
$T=0$\,K is set close to a minimum of any one of spin-split subbands.
When $\tau_0$ is reduced from $10^3$ (red curve) to $10$ (blue curve) for softer
potential edges, $G$ is only slightly decreased at low $T$, but is significantly
increased at high $T$.
We also find, from the upper-right panel of Fig.\,\ref{FIG:7}, that $S_{\rm d}$
increases with $T$ but decreases with $n_{\rm 1D}$. The decrease of $S_{\rm d}$
with $n_{\rm 1D}$ is simply due to the increase of $G$. As expected, $S_{\rm d}$
varies linearly with $T$ for small values of $T$ but deviates from the linear
behavior for large values of $T$, as shown in the lower-right panel of
Fig.\,\ref{FIG:7}.\,\cite{huang1} However, $S_{\rm d}$, in this case, goes
towards its minimum value about zero at $n_{\rm 1D}=4\times 10^6$\,cm$^{-1}$ as
$T\to 0$ because $G$ approaches the quantized value $6\,e^2/h$ for this electron
density. When $\tau_0$ is reduced from $10^3$ (red curve) to $10$ (blue curve),
$S_{\rm d}$ is enhanced at high $T$.
\medskip

\begin{figure}[p]
\includegraphics[width=7.0in,height=3.0in]{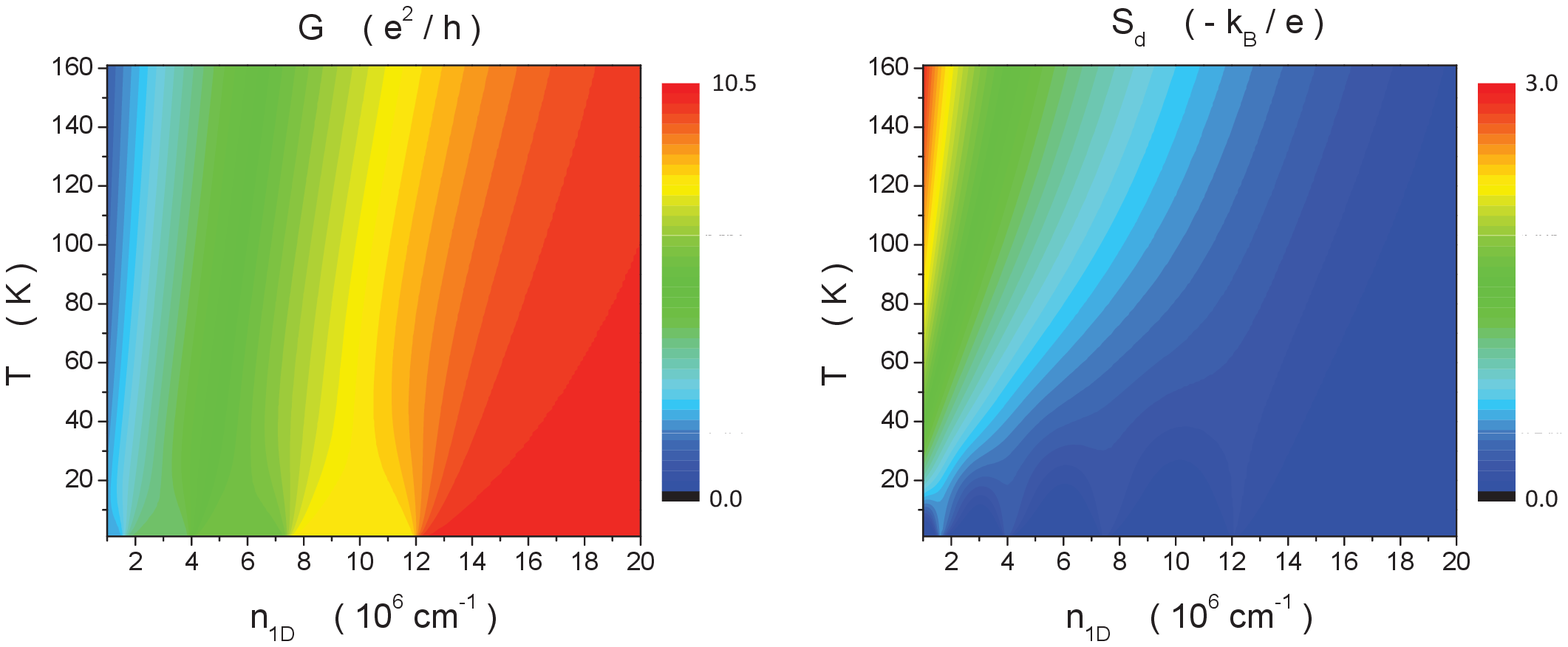}
\includegraphics[width=3.5in,height=2.5in]{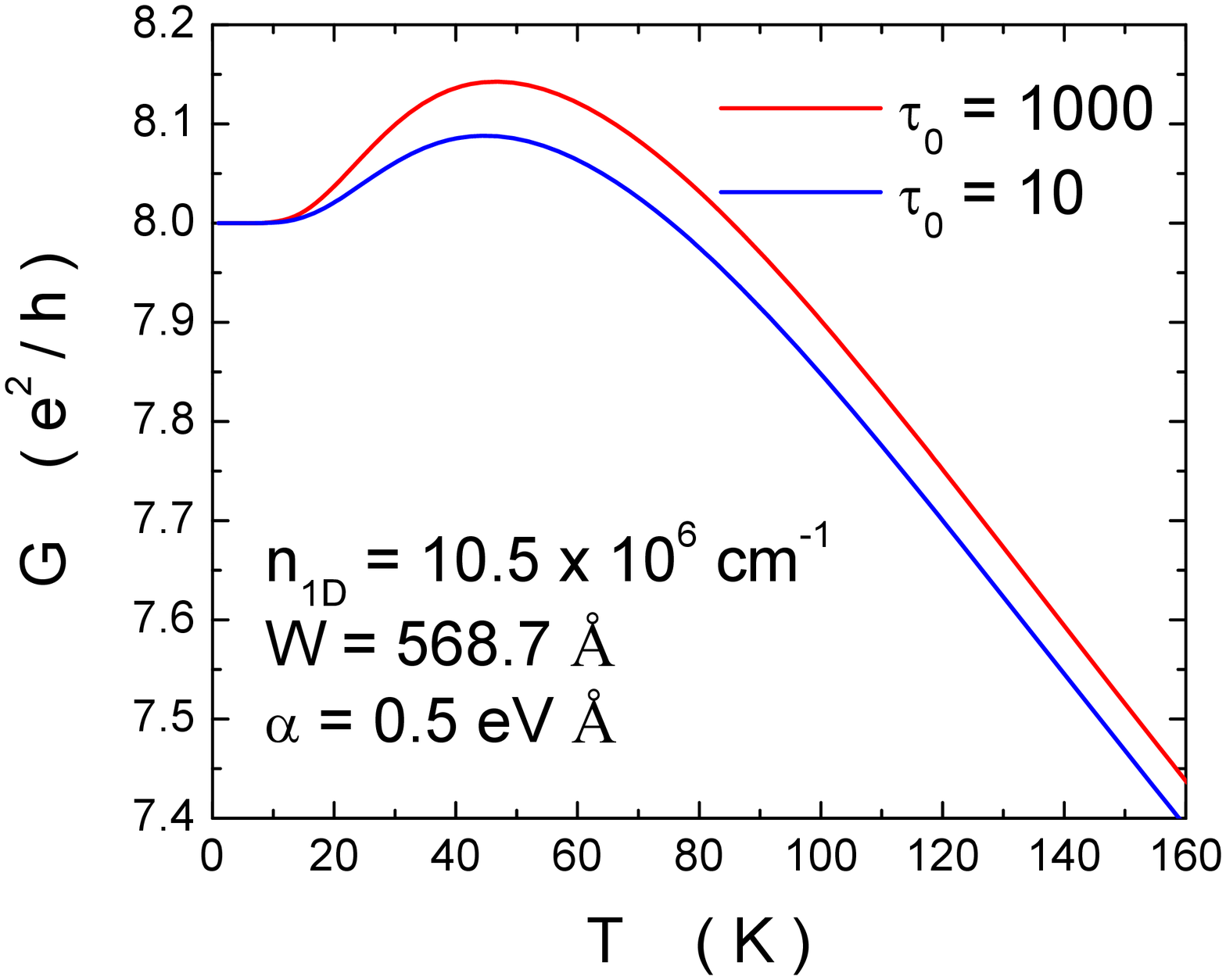}
\includegraphics[width=3.5in,height=2.5in]{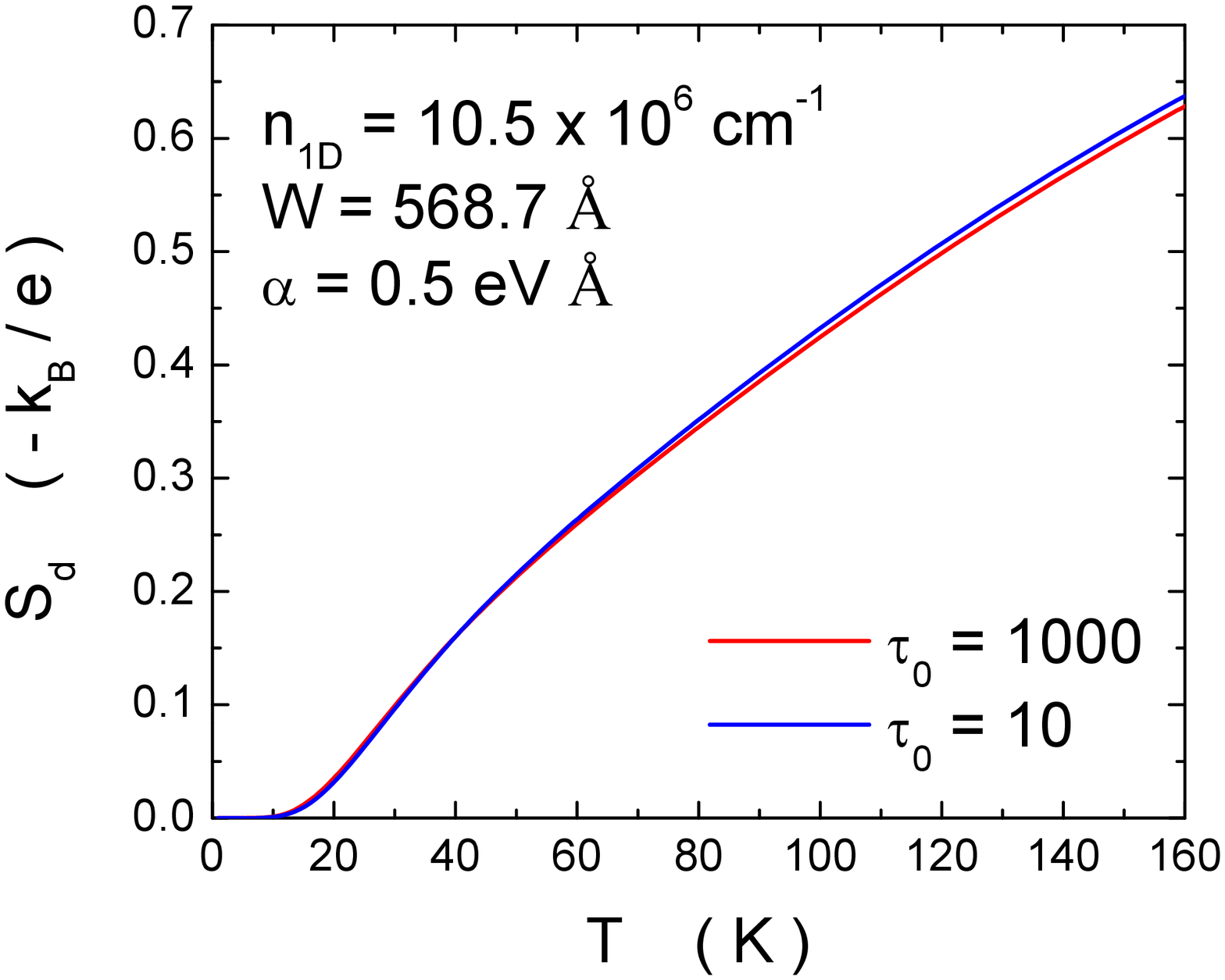}
\caption{(Color online) Contour plots of $G$ (upper left) and $S_{\rm d}$ (upper
right)
as functions of $T$ and $n_{\rm 1D}$ with $\alpha=0.5$\,eV$\cdot$\AA\ and ${\cal
W}=568.7$\,\AA. Here, $\tau_0=10^3$ is chosen for the upper two Contour plots.
For a fixed value of $n_{\rm 1D}=10.5\times 10^6$\,cm$^{-1}$, the plots for $G$
(lower left) and $S_{\rm d}$ (lower right) are also shown in this figure as a
function of $T$ with $\tau_0=10^3$ (red curves) and $\tau_0=10$ (blue curves).}
\label{FIG:8}
\end{figure}

Finally, we show another pair of contour plots for $G$ and $S_{\rm d}$ in
Fig.\,\ref{FIG:8} as functions of $T$ and $n_{\rm 1D}$ with ${\cal
W}=568.7$\,\AA\ and $\alpha=0.5$\,eV$\cdot$\AA. Here, similar to the two upper
panels of Fig.\,\ref{FIG:7}, $G$ (upper-left panel) increases with $n_{\rm 1D}$
but with much steeper steps, and $S_{\rm d}$ (upper-right panel) decreases with
$n_{\rm 1D}$ but at a more rapid rate at a higher temperature. At a higher
electron density $n_{\rm 1D}=10.5\times 10^6$\,cm$^{-1}$ in the lower-left panel
of Fig.\,\ref{FIG:8}, for a reduced value of ${\cal W}$, the range for anomalous
increase of $G$ with $T$ expands up to $40$\,K. This is further accompanied by a
locking of $G$ to its quantized value at $8\,e^2/h$ for $T\leq 10$\,K.
However, $G$ is reduced at high $T$ in this case when $\tau_0$ is reduced from
$10^3$ (red curve) to $10$ (blue curve).
Moreover, the locked $G$ value leads to an almost zero value for $S_{\rm d}$
within this temperature range, as seen from the lower-right panel of
Fig.\,\ref{FIG:8}. A clear linear dependence of $S_{\rm d}$ on $T$ is found for
$T>40$\,K due to suppressed thermal-population effect by a reduced wire width.
When $\tau_0$ is reduced from $10^3$ (red curve) to $10$ (blue curve), no
significant changes to $S_{\rm d}$ can be observed here.

\section{Concluding Remarks}
\label{sec5}

In this paper we investigated the effect that the
spin-orbit interaction has on the energy band structure, the conductance
and the electron-diffusion thermoelectric power of a nanowire. We used a model
in which
edge effects for the nanowires are taken into account by solving numerically
Dirac's
equation in a quasi-square potential. Comparing our model
with the already published work where harmonic confinement was used to
describe the transverse confinement, we found that the energy bands are
different and in addition to crossing effect of  the transverse energy bands,
there is also anticrossing for specific finite values of $k_y {\cal W}$.
The $\beta$-term of the Hamiltonian causes a
displacement and a deformation of the transverse energy band structure which is
more
pronounced for large values of $k_y {\cal W}$.
The conductance plateaus become wider when the  electron density is increased
as a result of larger energy-level separation as the higher subbands are
occupied by electrons.
Also, due to the nonlinear dependence of $\beta$-SOI on position close to the
well center,
there is no ``pole-like feature" in $G$ which is obtained when a harmonic
potential is
assumed for confinement. The electron-diffusion
thermoelectric power $S_{\rm d}$ displays a peak whenever a spin-split subband
is populated.
At low temperature, the variation of $S_{\rm d}$ is linear in $T$
but deviates from the linear behavior for large values of $T$.

We note that if the effect due to electron interaction is strong in our
 system, the ballistic model  in our paper cannot be justified.
For high-mobility semiconductor quantum wires, electron transport
is expected to be ballistic if the wire length is shorter than the
 mean free path of electrons in the system. On the other hand,
electron transport can also be diffusive if the wire length is
longer than the mean free path of electrons but less than the
 localization length. In the latter case, the interaction of
 electrons with impurities, roughness, phonons and other electrons
will play a significant role in both the temperature and density
dependence of electron mobility in quantum wires. For high temperature
and low density, electron-phonon scattering is dominant.
However, the electron-electron scattering becomes significant in
 a system with high density at low temperature.  In this
 paper, we restricted ourselves to the ballistic regime for a short
 quantum wire.

Furthermore, In the  absence of electron-electron interaction,
an effect of lateral spin-orbit coupling  on transport
is to trigger a spontaneous but negligible spin polarization
 in the nanowire. However, the spin polarization
is enhanced substantially when the effect of electron-electron
 interaction is included. The spin polarization may be
strong enough to result in the appearance of a conductance
plateau at a fractional value  ($\sim 0.7$) of  $2e^2/h$  in the absence of
 any external magnetic field.\cite{pepper1,pepper2,thomas1,thomas2}
The role played by electron-electron interaction on our results for
the thermoelectric power in a nanowire  may also be investigated
using finite-temperature Green's  function techniques or field
theoretic methods which would allow the classification of contributing
diagrams.\cite{electron-electron} This study should shed some light
on whether electron-electron interaction enhances the thermoelectric power
and  how the SOI underlies the Peltier effect, i.e., the flow of
entropy current in addition to the familiar charge current in
an electric field. The effect of electron-electron interaction is
 expected to be significant at very low temperatures (less than 1K)
 and high electron densities. However, our model calculations
 are performed at temperatures at or higher than 4K.

\acknowledgments

This research was supported by contract \# FA 9453-07-C-0207 of AFRL. DH would
like
 to thank the Air Force Office of Scientific Research
(AFOSR) for its support.

\newpage

\begin{thebibliography}{99}
\bibitem{davies} J. H. Davies, {\em The Physics of Low-Dimensional
Semiconductors},
(Cambridge University Press, New York, 1998).

\bibitem{pepper1}L. W. Smith, W. K. Hew, K. J. Thomas, M. Pepper, I. Farrer, D.
Anderson,
G. A. C. Jones, and D. A. Ritchie, \prb {\bf 80}, 041306 (2009).

\bibitem{pepper2}W. K. Hew, K. J. Thomas, M. Pepper, I. Farrer, D. Anderson, G.
A. C. Jones,
and D. A. Ritchie, \prl {\bf 102}, 056804 (2009).

\bibitem{moroz}A. V. Moroz and C. H. W. Barnes,
\prb {\bf 60}, 14272 (1999).

\bibitem{privman}Y. V. Pershin, J. A. Nesteroff, and V. Privman,
\prb {\bf 69}, 121306 (2004).

\bibitem{gumbs1}G. Gumbs, \prb  {\bf 70}, 235314 (2004).

\bibitem{manvir}M. S. Kushwaha, \prb {\bf 76}, 245315 (2007).

\bibitem{gumbs}G. Gumbs, \prb {\bf 73}, 165315 (2006).

\bibitem{go}G.-Q. Hai and M. R. S. Tavares,
\prb {\bf 61}, 1704 (2000).

\bibitem{fertig}L. Brey and H. Fertig, \prb {\bf 75}, 125434 (2007).

\bibitem{darwin}C. G. Darwin,  Proc. Roy. Soc. (London) {\bf 120}, 621 (1928).

\bibitem{fisher}G. P. Fisher,  Am. J. Phys. {\bf 39}, 1528 (1971).

\bibitem{foldy}L. L. Foldy and S. A. Wouthuysen, Phys. Rev. {\bf 78}, 29 (1950).

\bibitem{Beenakker}C. W. J. Beenakker and H. van Houten,
{\em Quantum Transport in Semiconductor Nanostructures}, Vol. {\bf 44} of Solid
State Physics (Academic Press, New York, 1991).

\bibitem{LYO:2004} S. K. Lyo and D. H. Huang, J. Phys.: Condens. Matter, {\bf
16}, 3379 (2004).

\bibitem{num12} C. Itzykson and J.-B. Zuber, {\em Quantum Field Theory\/}
(cGraw-Hill, New York, 1980).

\bibitem{num13} V.K. Thankappan, {\em Quantum Mechanics\/}, (John Wiley and
Sons, New York, 1993).

\bibitem{Rashba-paper} Yu. A. Bychkov and E. I. Rashba, J. Phys. C {\bf 17},
6039 (1984).

\bibitem{num19} M. F. Li, {\em Modern Semiconductor Quantum Physics\/},
(World Scientific, Singapore, 1994).

\bibitem{num20} B. K. Ridley, {\em Quantum Processes in Semiconductors\/}
(Clarendon
Press, Oxford, 1993).

\bibitem{referee} T. Darnhofer and U. R�ssler, Phys. Rev. B{\bf 47}, 16020
(1993).

\bibitem{huang1}S. K. Lyo and D. H. huang, \prb {\bf 66}, 155307 (2002).

\bibitem{thomas1}   K. J. Thomas, J. T. Nicholls, M. Y. Simmons, M. Pepper,
 D. R. Mace, and D. A. Ritchie, Phys. Rev. Lett. {\bf 77}, 135 (1996).

\bibitem{thomas2} K. J. Thomas, J. T. Nicholls, N. J. Appleyard, M. Y. Simmons,
M. Pepper, D. R. Mace, W. R. Tribe, and D. A. Ritchie , Phys. Rev. B {\bf 58},
4846 (1998).

\bibitem{electron-electron} J. W. P. Hsu, A. Kapitulnik, and M. Yu. Reizer,
 Phys. Rev. B {\bf 40}, 7513  (1989).


\end{thebibliography}
\end{document}